\documentstyle[12pt,eqsecnum,aps,epsf]{revtex}
\tighten

\def\dirac{{\rm D}\!\!\!\!/\,}

\def\ham{{\rm H}}

\def\bmat{{\rm B}}
\def\cmat{{\rm C}}

\def\lpmb#1{\mbox{\boldmath$#1$}}

\begin{document}

\begin{center}
\LARGE{\bf Spontaneous Flavor and Parity Breaking \\
		with  Wilson  Fermions}
\end{center}

\vskip0.5cm
\centerline{Stephen Sharpe\footnote[2]
{sharpe@phys.washington.edu}
 and Robert Singleton, Jr.\footnote[3]
{bobs@terrapin.phys.washington.edu}}
\smallskip
\centerline{\it Physics Department}
\centerline{\it Box 351560, University of Washington}
\centerline{\it Seattle, WA  98195, USA}

\setcounter{page}{0}
\thispagestyle{empty}

\vskip 1in

\centerline{\bf Abstract}
\vskip 0.2in

\begin{quote}
We discuss the phase diagram of Wilson fermions in the
$m_0$--$g^2$ plane for two-flavor QCD. We argue that, as
originally suggested by Aoki, there is a phase in which flavor
and parity are spontaneously broken. Recent numerical results on
the spectrum of the overlap Hamiltonian have been interpreted as
evidence against Aoki's conjecture. We show that they are in fact
consistent with the presence of a flavor-parity broken ``Aoki
phase''. We also show how, as the continuum limit is approached,
one can study the lattice theory using the continuum chiral
Lagrangian supplemented by additional terms proportional to
powers of the lattice spacing. We find that there are two
possible phase structures at non-zero lattice spacing: (1) 
there is an Aoki phase of width $\Delta m_0 \sim a^3$ with two
massless Goldstone pions; (2) there is no symmetry breaking, 
and all three pions have an equal non-vanishing mass of order
$a$. Present numerical evidence suggests that the former option
is realized for Wilson fermions. Our analysis then predicts the
form of the pion masses and the flavor-parity breaking condensate
within the Aoki phase. Our analysis also applies for
non-perturbatively improved Wilson fermions.
\end{quote}

\vfill

\noindent UW/PT 98-2

\noindent hep-lat/9804028 \hfill Typeset in La\TeX

\eject

\section{Introduction}
\label{sec:intro}

\begin{figure}
\centerline{
\epsfxsize=85mm
\epsfbox{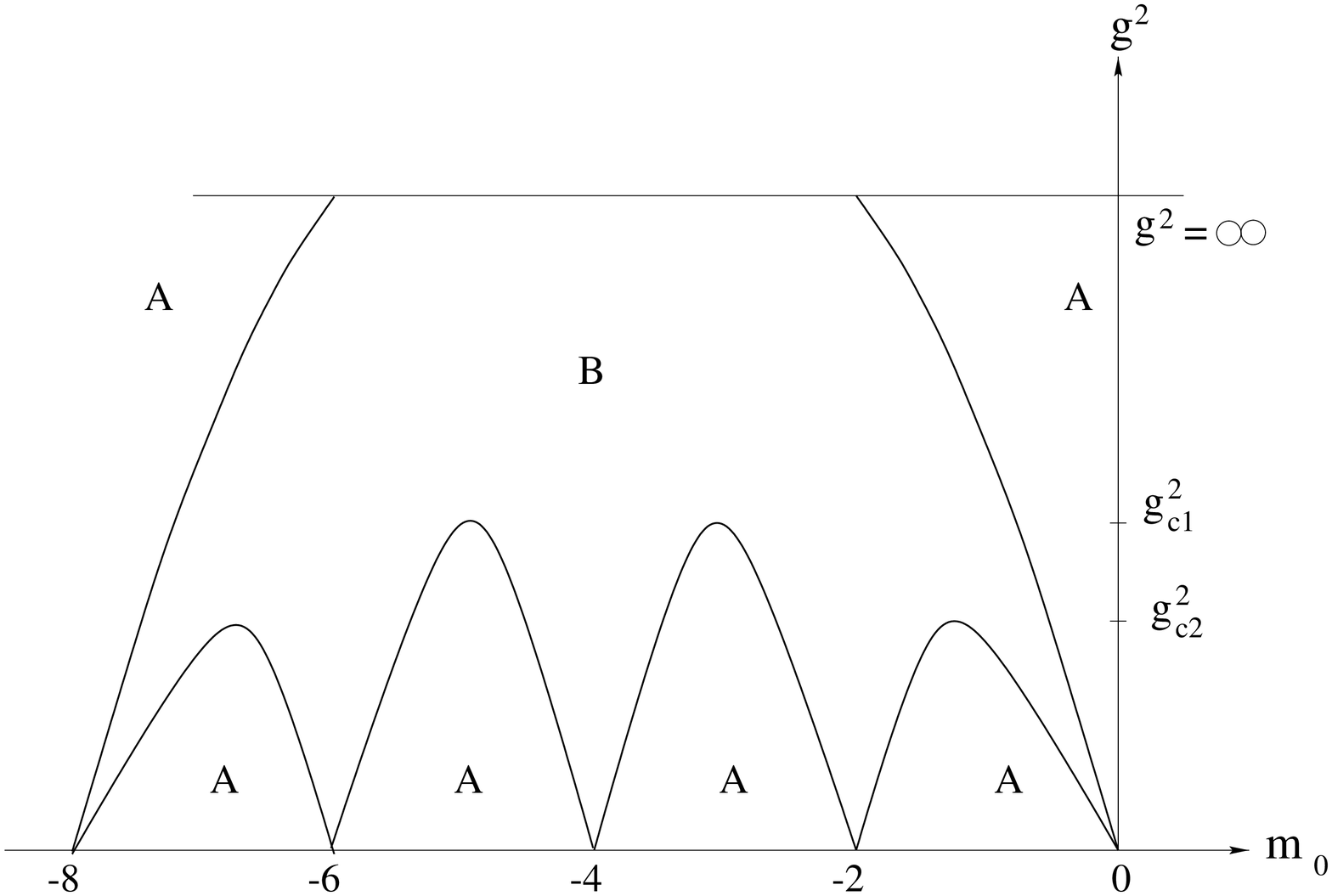}
}
\vskip0.5cm
\caption{The phase diagram proposed by Aoki: $g$ is the gauge
coupling and $m_0$ the dimensionless bare quark-mass. 
The continuum-like phases are labeled A, 
and the flavor and parity broken phase B. 
The phase diagram is symmetric under $m_0\leftrightarrow -(m_0+8)$.
The continuum limit of particular interest is that at $m_0=0$, $g=0$.}
\label{fig:aoki}
\end{figure}

This paper concerns the phase diagram of Wilson fermions (and
improved versions thereof) at non-zero lattice spacing, and the
restoration of chiral symmetry in the continuum limit. Some time
ago, Aoki proposed a phase diagram in which there were regions of
spontaneous flavor and parity breaking (which we refer to as Aoki
phases)\cite{aokiphase}. This suggestion, sketched in
Fig.~\ref{fig:aoki}, provides a dynamical explanation for the
masslessness of the pion on the lattice. Analytical and numerical
support for this proposal has been given in
Refs.~\cite{aokiphase,aokigocksch,akuu,aku,bitarphase}.

The validity of this picture has, however,
been challenged recently by Bitar, Heller and Narayanan\cite{BHN}.
These authors argue that flavor and parity violation do not occur
at non-zero lattice spacing, and that, by analogy with the
massless continuum theory, one should interpret the Aoki 
phases as containing massless quarks and exhibiting spontaneous 
chiral symmetry breaking. Numerical results supporting this 
claim are given in Refs.~\cite{EHNS,EHNII}. 

In this paper we present two arguments supporting Aoki's
suggestion of spontaneous flavor and parity breaking at non-zero
lattice spacing. Our first observation is that the numerical
results of Refs.~\cite{EHNS,EHNII}
are not only consistent with Aoki's picture, but they are 
inconsistent with the interpretation of Bitar {\em et al.}.
Thus all numerical results to date are consistent with
Aoki's proposed symmetry breaking.

Our second and more important observation is that, close to the
continuum limit, one can study the pattern of symmetry breaking
theoretically, using the chiral Lagrangian. One must augment the
usual continuum chiral Lagrangian with terms corresponding to
the explicit breaking of chiral symmetry at non-zero lattice
spacing. The form of these terms is dictated by the symmetries of
the lattice theory. Their coefficients are, however, undetermined
aside from their order of magnitude. Nevertheless, it turns out
that this is sufficient information to determine the pattern of
symmetry breaking at small but non-zero lattice spacing, up to a
two-fold ambiguity. If the sign of a particular coefficient is
positive, then there is an Aoki phase with all the expected
properties. If the sign is negative, then there is no Aoki-phase,
flavor is always unbroken, and the pions do not become massless
for any value of the bare mass. The present numerical evidence
suggests that the first option correctly describes (unimproved)
Wilson fermions. If this is the case, our calculation makes
several predictions that can be tested by numerical simulations.

In Aoki's picture, the flavor-parity violating phases shrink 
rapidly to isolated points as one approaches the continuum limit
(see Fig.~\ref{fig:aoki}).
At these points, the quarks are massless, chiral-symmetry
is spontaneously broken, and all pions are massless. 
Our analysis finds that the width of the Aoki phase shrinks
as $\Delta m_0\propto a^3$, up to logarithmic corrections.
Perhaps more importantly, we can see how the
flavor-parity breaking at non-zero lattice spacing transforms 
smoothly into chiral symmetry breaking in the continuum limit.
In our view, the interpretation of Bitar {\em et al.} 
applies only in this limit. 

Similar observations concerning symmetry breaking have been made 
previously by Creutz~\cite{creutz}.
He performs a qualitative analysis using the linear sigma-model,
and finds the same two possible patterns of symmetry
breaking at finite lattice spacing.
Our calculation extends his by providing
a firmer theoretical basis for his observations
and by making quantitative predictions.

The outline of this paper is as follows. In the next section we
review Aoki's proposal for the phase diagram, and then describe
the alternative view of Bitar {\em et al.}. In
Sec.~\ref{sec:cons} we explain how the numerical results obtained
to date are all consistent with Aoki's proposal. We then present,
in Sec.~\ref{sec:chiral}, our Chiral Lagrangian analysis.
Sec.~\ref{sec:conc} contains our conclusions. We collect
technical issues in two appendices.

\section{Review of Previous Results}
\label{sec:review}

We consider QCD with two degenerate quarks, for which the
fermionic part of the Euclidean lattice action is
\begin{eqnarray}
  {\cal L}_f =  \bar\psi W(m_0) \psi \ ,
\label{twoflath}
\end{eqnarray}
where the flavor indices are implicit, and $W(m_0)$ is the
Wilson-Dirac operator with a (dimensionless) bare mass $m_0$
common to both flavors. The bare mass is related to the usual
hopping parameter by \hbox{$m_0 = {1/(2\kappa)}-4$}. The action
is invariant under $SU(2)$ flavor transformations, under parity,
and under the interchange $m_0\to -(m_0+8)$, but it explicitly
breaks all axial symmetries for any value of $m_0$.

We are concerned here with the phase diagram in the $m_0$--$g^2$
plane, where $g$ is the gauge coupling. Based on perturbation
theory, we expect to be able to take a continuum limit by
simultaneously sending $g^2\to 0$ and $m_0\to0$. One can, however,
obtain different continuum limits by sending $m_0$ to any one
of the values $0$, $-2$, $-4$, $-6$, or $-8$, for which there
are respectively $1$, $4$, $6$, $4$ and $1$ continuum fermions. 
In the following we will mainly be interested in the standard 
continuum limit at $m_0=0$ (or the equivalent point at $m_0=-8$).
If this point is approached
in such a way that the physical quark-mass is small compared to 
$\Lambda_{\rm QCD}$, then we expect the theory to spontaneously
break chiral symmetry and for there to be a degenerate triplet 
of light pions. This should be true whether we approach from
above or below $m_0=0$, since in the continuum the sign of the
quark mass is irrelevant. Similar phenomena should occur at the
other continuum points, except that the symmetry group in the
continuum limit will be larger ($SU(8)$ at $m_0=-2$ and $-6$, and
$SU(12)$ at $m_0=-4$), and so there should be correspondingly
more pions.\footnote{At non-zero lattice spacing, however, only
an $SU(2)$ subgroup of the full continuum flavor symmetry is
exact, so the full pion multiplet will be broken into
representations of this subgroup, with the lightest pions being
in a triplet.} In fact, it may be that the dynamics with these
large numbers of flavors is different from the $m_0=0$ and $-8$
cases, since above a critical number of flavors we expect there
to be no chiral symmetry breaking and no
confinement~\cite{bankszacs}.

We are particularly interested in how the phenomena associated
with chiral symmetry breaking emerge in the continuum limit,
given that the lattice theory has no such symmetry. In
particular, if the continuum limit is approached so that the
quarks are massless, then the resulting theory should contain
massless Goldstone pions. The standard view of how the theory
accomplishes this is that there is a critical line, $m_c(g^2)$,
along which the triplet of pions is massless. This line runs from
$m_0=-2$ at $g^2=\infty$ to the continuum point at $m_0=0$, $g^2=0$.
At non-zero $g^2$, the pions are not Goldstone particles
associated with spontaneous chiral symmetry breaking (and in
particular their interactions are not restricted by chiral
symmetry), but they become so in the continuum limit.

\subsection{Aoki's proposal}
\label{sec:aoki}

Based on a variety of considerations, Aoki proposed that the
vanishing of the pion mass is associated with the spontaneous
breakdown of flavor symmetry~\cite{aokiphase}. His proposal for
the phase
diagram is sketched in Fig.~\ref{fig:aoki}. The phases labeled A
are ``continuum-like'' in the sense that the $SU(2)$ flavor
symmetry is unbroken. In phase B, by contrast, flavor symmetry is
spontaneously broken down to $U(1)$, and parity is also
spontaneously broken. There are thus two Goldstone bosons in this
phase, which are exactly massless in infinite volume, even though
the lattice spacing is non-zero. We will call these
the charged pions. The neutral pion which completes the
flavor triplet is not a Goldstone particle, and is thus massive
within phase B. The transitions between the two phases are
continuous, so that as one approaches the boundary from within
phase B, the neutral pion must also become massless so as to
restore flavor symmetry. This means that as one approaches the
boundary from within phase A, the degenerate triplet of pions
becomes massless. The critical line discussed above is the
rightmost phase boundary.

This proposal has been established at strong
coupling\cite{aokiphase}, for which there are just two critical
lines (related by the symmetry $m_0\to -(m_0+8)$). At weak coupling,
if one is to reproduce the continuum limits expected from
perturbation theory, then one must introduce the additional
regions of phase A shown in the figure\footnote{%
These new regions of phase A appear below critical values of $g^2$, 
denoted $g_{c1}^2$ and $g_{c2}^2$ in the figure. These are
conventionally drawn as being equal, but we can see no reason to
expect this equality. The particular ordering of $g_{c1}^2$ and
$g_{c2}^2$ shown in the figure is an arbitrary choice.}.
In particular, if the continuum limit at $m_0=0$ can be approached
from positive and negative mass, then the width of phase B must
vanish at $g^2=0$. Aoki argues that, since phase B is not
observed in perturbation theory, its width must be a
non-perturbative function of $g^2$, {\em i.e.} have a power law
dependence on $a$. We argue in Sec.~\ref{sec:chiral}
that the dependence is $\Delta m_0\propto a^3$ (up to
logarithms).

To investigate the spontaneous symmetry breaking in phase B, 
Aoki added to the Lagrangian (\ref{twoflath}) a source term
that explicitly breaks both flavor and parity, 
\begin{equation}
  {\cal L}_{\rm source} = i h \bar\psi \gamma_5 \sigma_3 \psi \,,
\label{source1}
\end{equation}
with $\sigma_3$ acting in flavor space. Flavor and parity are
spontaneously broken if, in the limit that the source is removed,
a condensate remains:
\begin{equation}
  \lim_{h\to 0^{\pm}}
  \langle i \bar\psi \gamma_5 \sigma_3 \psi \rangle = 
  \pm {\rm const} \,. 
\label{condlat}
\end{equation}
Flavor rotations in the 1- and 2-directions change the condensate,
while those in the 3-direction do not. The symmetry is thus
broken from $SU(2)\to U(1)$, with the condensates inhabiting 
the coset space $SU(2)/U(1)$. The condensate (\ref{condlat})
appears to violate theorems by Vafa and Witten which state that
vector-like theories do not spontaneously break flavor\cite{vw1}
or parity\cite{vw2}. As explained in Appendix~\ref{app:vw}, however,
these theorems are not applicable to the case at hand.

Numerical evidence supporting Aoki's proposal has been given in
Refs.~\cite{aokiphase,aokigocksch,akuu,aku,bitarphase}. In
particular,  Ref.~\cite{aku} studied the pion mass as a
function of $m_0$ at $g^2=1$ ($\beta=6/g^2=6$) in the quenched
approximation, finding evidence for all 10 critical lines
predicted near the continuum limit (although this required the
removal by hand of so called ``exceptional configurations'').
Unquenched simulations in small volumes also found evidence of
Aoki phases, although Ref.~\cite{bitarphase} did not find them at
the smallest couplings studied.
This could, however, be due to the small width of the phases.

\subsection{The proposal of Bitar, Heller and Narayanan}
\label{sec:bitar}

The phase diagram for Wilson fermions has recently been
reconsidered in Refs.~\cite{BHN,EHNS,EHNII}. In Ref.~\cite{BHN},
Bitar {\em et al.} take issue with Aoki's proposal. They point
out that, in the massless limit of the continuum theory, a
condensate of the form of (\ref{condlat}) is simply an
axial rotation of the usual flavor diagonal condensate, and thus
breaks neither flavor nor parity. In other words, the condensate
(\ref{condlat}) spontaneously breaks the chiral group
$SU(2)_L\times SU(2)_R$ down to an $SU(2)$ subgroup, albeit a
different subgroup from the usual
flavor group (its generators are vector rotations in the
3-direction and axial rotations in the 1- and 2- directions).
This unbroken subgroup can be defined to be the flavor symmetry.
Similarly, the condensate preserves a discrete symmetry, which is 
the original parity transformation conjugated by an axial
rotation, and this combined symmetry can be taken to be parity.
Based on these observations, they argue that the Aoki phase
should be interpreted as having massless quarks and spontaneous
chiral symmetry breaking. The condensate of
(\ref{condlat}) should be interpreted as (an axial
rotation of) the usual chiral condensate. Flavor and parity
symmetry, they claim, are not broken at non-zero lattice spacing.

To study the issue numerically, Refs.~\cite{EHNS,EHNII}
investigate the spectrum of the overlap Hamiltonian
$H(-m_0)=\gamma_5 W(m_0)$.\footnote{The sign choice of $m_0$ is 
the convention introduced in Refs.~\cite{ol} and adopted 
also in Refs.~\cite{BHN,EHNS,EHNII}.} The advantage of this
operator over the Wilson-Dirac operator is that it is Hermitian, 
and so its eigenvalues are confined to the real axis.
This allows one to define a spectral density%
\footnote{Spectral densities can be defined on a given gauge
configuration by taking the infinite volume limit. We refer to
these as $\rho_A(\lambda;m_0)$ in the continuum, and
$\rho_U(\lambda;m_0)$ on the lattice, with the proviso that on
the lattice we average over a configuration and its parity conjugate.
The quantity of interest here, $\rho(\lambda;m_0)$, is obtained by 
averaging over all configurations, with the appropriate weights.
For further discussion see Appendix~\ref{app:calcs}.}
$\rho(\lambda;m_0)$ both on the lattice and in the continuum. As
explained in Refs.~\cite{EHNS,EHNII}, the central issue is
whether this spectral density, considered as a function of
$\lambda$ for fixed $m_0$, vanishes for a region around
$\lambda=0$. In other words, does the spectrum have a gap?
This is important because, as we now explain, the absence of a
gap is indicative of chiral symmetry breaking in the continuum.

We begin by recalling that the Dirac operator in the continuum,
unlike the Wilson-Dirac operator on the lattice, has a spectrum
confined to the imaginary axis. Thus one can define a spectral
density $\tilde\rho(\lambda)$ for the operator $-i\dirac$. As
first noted by Banks and Casher\cite{BC}, chiral symmetry
breaking occurs if $\tilde\rho(\lambda)$ does not vanish at
$\lambda=0$. In the continuum, the spectrum of the overlap
Hamiltonian, which is just $\gamma_5(\dirac + m)$, is related to
that of $-i\dirac$ by
\begin{equation}
  \rho(\lambda;m) = 
  \left\{ \begin{array}{cc}
  \frac{\textstyle |\lambda|}{\sqrt{\textstyle \lambda^2-m^2}} \,
  \tilde\rho(\sqrt{\lambda^2-m^2}) & ~~~~|\lambda|>|m| \\
  ~ & ~ \\
  0 & ~~~~|\lambda| \le |m| \end{array} \right. \ .
\label{contsd}
\end{equation}
We see that at non-zero quark-mass, $\rho(\lambda;m)$ has a gap
between $\lambda=-|m|$ and $|m|$. As the quark mass $m\to0$, 
note that $\rho(\lambda;m)$ converges to $\tilde\rho(\lambda)$,
albeit non-uniformly. Thus the gap closes in the chiral limit
and $\rho(\lambda;0)=\tilde\rho(|\lambda|)$. Based on this
continuum result, Ref.~\cite{BHN} argues that, if the spectrum of
$\gamma_5 W(m_0)$ has no gap, then one should interpret the
lattice theory as having massless quarks and chiral symmetry
breaking.

As noted in Ref.~\cite{BHN}, the spectral density at zero
eigenvalue is related to the condensate (\ref{condlat}) proposed
by Aoki,
\begin{equation}
  \lim_{h\to 0^{\pm}}
  \langle i \bar\psi \gamma_5 \sigma_3 \psi \rangle
	=\mp 2\pi \rho(0;m_0) \,.
\label{condlatrho}
\end{equation}
Thus Bitar {\em et al.} do expect that the Aoki phases should be
characterized by a condensate, as well as vanishing pion masses.

References~\cite{EHNS,EHNII} study
the low-lying spectrum of $\gamma_5 W(m_0)$ 
for the mass range $-2 \le m_0\le 0$.
They have results on a variety of gauge field ensembles
(quenched with and without improvement, and partially quenched)
all with $g^2 \approx 1$.
They find that in all cases the gap in the spectrum closes for
bare masses below a critical line, {\em i.e.} for $m_0 < m_1(g^2)$.
This lies very close to the critical curve $m_c(g^2)$ determined
from the vanishing of the extrapolated pion mass. 
We follow Ref.~\cite{EHNII} and assume that the curves would 
coincide were one to extrapolate including the effects of quenched chiral
logarithms. What is perhaps unexpected, however, is that the gap does not
open up before they reach $m_0=-2$.
This is inconsistent with the
results of Ref.~\cite{aku}, who find that the Aoki phase
has a width of $\Delta m_0\approx 0.1$.

\section{Consistency}
\label{sec:cons}

In this section we address the conflict between the
interpretations of Bitar {\em et al.} and Aoki. We begin with
points of agreement. Both contend that there are two phases,
and that phase B is characterized by a condensate of the form
(\ref{condlat}). They also agree that the masses of all three
pions vanish along the boundaries between phases. The conflict
concerns the number of massless pions within phase B and the
interpretation of the condensate.

We think that Aoki's proposal is a correct description of what
happens at any non-zero lattice spacing, while the interpretation
of Bitar {\em et al.} is an approximate description that becomes
valid in the continuum limit. In this section we elaborate on the
former point, while in the next we present a framework within
which to understand the latter.

The essential point has already been made by Aoki\cite{aokiphase}.
The only exact symmetry of the lattice theory is the $SU(2)$
flavor symmetry, and the condensate (\ref{condlat}) spontaneously
breaks this symmetry down to $U(1)$. Thus Goldstone's theorem
implies that there are only two massless pions in the broken
phase.

It is worthwhile discussing the symmetry breaking in more detail, 
so as to emphasize the differences between the massless continuum
theory and the lattice in phase B. To study spontaneous symmetry
breaking, one adds a source term that explicitly breaks the
symmetry, calculates the order parameter (here the condensate),
and then takes the limit of vanishing source strength. If the
symmetry is spontaneously broken from $G\to H$, the resulting
condensate lives in the coset space $G/H$. One can thus explore
the coset space by examining the condensate for sources that
effect breaking to different $H$-subgroups of $G$. This leads 
us to consider, both on the lattice and in the continuum, the
following source:
\begin{equation}
  {\cal L}_{\rm source}({\lpmb \theta}, h) = 
  h \bar\psi \exp\left[i {\lpmb \theta}
  \cdot {\lpmb \sigma}\gamma_5\right] \psi \ ,
\label{source2}
\end{equation}
where ${\lpmb \sigma}$ is a three-vector of Pauli matrices that
acts only in flavor space. This source is obtained by applying
a general chiral transformation to the mass term $h \bar\psi\psi$.
We will denote the norm of ${\lpmb\theta}$ by $\theta$, while
$\hat{\lpmb\theta}$ will represent a unit vector in the direction
of ${\lpmb\theta}$. Without loss of generality we can take $h >
0$ and $0\le \theta < \pi$ (so that $\sin\theta \ge 0$). In the
massless continuum theory, standard arguments presented in
Appendix~\ref{app:calcs} lead to
\begin{mathletters}
\label{condcont}
\begin{eqnarray}
  \lim_{h\to 0^+}\langle \bar\psi_b \psi_a 
  \rangle &=& -\pi\, \tilde\rho(0)
  \cos\theta\,\delta_{ab}
\label{condconta}\\
  \lim_{h\to 0^+}\langle \bar\psi_b i\gamma_5 \psi_a \rangle
  &=& -\pi\, \tilde\rho(0)\sin\theta \,
  [\hat{\lpmb \theta}   \cdot {\lpmb \sigma}]_{ab} \ .
\label{condcontb}
\end{eqnarray}
\end{mathletters}
where $a$ and $b$ are flavor indices, and $\tilde\rho(\lambda)
=\rho(\lambda;m=0)$ is the spectral density of the Dirac operator
introduced above. If chiral
symmetry is broken down to an $SU(2)$ subgroup, then the
condensates live in $SU(2)_L\times SU(2)_R/SU(2)$, and should be
described by three parameters. This is indeed what we find. 

The calculation in the lattice theory is somewhat different, 
and is explained in Appendix~\ref{app:calcs}. Only the
pseudoscalar condensate is proportional to the spectral density,
and we find
\begin{equation}
\label{nonfdiaglat}
  \lim_{h\to0^+}
  \langle \bar\psi_b i\gamma_5 \psi_a \rangle =
  \left\{ \begin{array}{cc}
  - \pi\, \rho(0;m_0)\, [\hat{\lpmb\theta}\cdot {\lpmb
  \sigma}]_{ab} & ~~~ \theta> 0 \\
  0 & ~~~ \theta=0 \end{array} \right. \,.
\label{condlattwo}
\end{equation}
Equation~(\ref{condlatrho}) is a particular case of this general
result. The essential difference between (\ref{condlattwo}) and
the corresponding continuum result (\ref{condcontb}) is the
absence of the factor $\sin\theta$. This means that the lattice
condensate is of fixed magnitude, and varies in direction
according to the applied source. It depends on only two
parameters, the angles in $\hat{\lpmb \theta}$, and maps out 
the coset space $SU(2)/U(1)$. This clearly represents a different
pattern of symmetry breaking than in the massless continuum
theory.

We wish to emphasize that these two different modes of
spontaneous symmetry breaking may be differentiated by simply
enumerating the number of massless pions:
in the massless continuum theory, there are three massless pions,
while on the lattice in phase B there are only two. To test this,
one needs to know which interpolating fields couple to the
different pions. For definiteness we consider a source with
$\theta=\pi/2$ in the 3-direction, {\em i.e.} ${\lpmb
\theta}=\frac{\pi}{2}\,\hat{\lpmb z}$, which leads to the
condensate
$ \langle \bar\psi i\gamma_5 \sigma_3\psi \rangle \ne 0$.
This is the choice used in the numerical work described in the
previous section. Interpolating fields for the Goldstone pions
are obtained by doing a spatially dependent infinitesimal $G/H$
transformation on the condensate. In the massless continuum
theory, these transformations are the vector transformations in
the 1- and 2-directions, $\psi(x)\longrightarrow \exp(-i\alpha(x)
\sigma_{1,2}) \psi(x)$ and the axial transformation in the
3-direction, $\psi(x)\longrightarrow \exp(-i\alpha(x) \sigma_3
\gamma_5)\psi(x)$. These yield the pion fields
\begin{equation}
  \pi_1(x) = i \bar\psi(x) \sigma_2 \gamma_5\psi(x) \,,\quad
  \pi_2(x) = -i \bar\psi(x) \sigma_1 \gamma_5\psi(x) \,,\quad
  \pi_3(x) = \bar\psi(x) \psi(x) \,.
\end{equation}
On the lattice axial transformations are not a symmetry, and
only the first two interpolating fields correspond to Goldstone
modes. We see that the charged pions are created by the familiar
pseudoscalar flavor non-singlet  operators, although with the
flavor indices $1$ and $2$ interchanged. The two-point functions
of such operators have no disconnected contributions, even in the
presence of the source term. These are the pions whose masses
have been calculated in numerical simulations.

The neutral pion, on the other hand, is created by an operator
that appears to be a scalar and a flavor singlet. This is
because the condensate itself points in a flavor and parity
violating direction. Indeed, when calculating the neutral pion
correlator it is crucial to do so with the source turned on, so
as to pick out the vacuum state, and then to take the limit of
vanishing source strength. The correlator for this operator has
not been studied numerically. We expect such a calculation will
be difficult, because it contains both disconnected and connected
contractions, 
\begin{equation}
  \langle \pi_3(x) \pi_3(y) \rangle_U = {\sf Tr} S_U(x;x) 
  {\sf Tr} S_U(y;y) - {\sf Tr} \left\{S_U(x;y) S_U(y;x) 
  \right\}\,.\end{equation}
Here $S_U(x;y)$ is the fermion propagator in the presence of the
source. While in the continuum, the disconnected contributions
can be shown to vanish because of chiral symmetry --- this must
be so since the three pions are degenerate --- this is not so at
non-zero lattice spacing. One might hope, however, that their
contribution will be small close to the continuum limit, and so
it would be interesting to calculate the connected contribution
alone. In the next section we give a prediction for the mass of
this particle in the Aoki phase.

\section{Chiral Lagrangian Analysis}
\label{sec:chiral}

In this section we study the extent to which the spontaneous
breaking of flavor and parity at non-zero lattice spacing can be 
viewed as an approximation to the spontaneous breaking of chiral
symmetry in the continuum limit. We do so by first determining
the effective continuum Lagrangian which describes the lattice
theory at non-zero lattice spacing, and then extracting the
effective chiral Lagrangian which describes the long distance
behavior of this effective continuum theory. The analysis relies
solely on symmetries and the assumption that dimensionful
quantities have a size determined by $\Lambda_{\rm QCD}$ to the
appropriate power. In the following we shall, for the
sake of brevity, drop the subscript on $\Lambda_{\rm QCD}$.

Close to the continuum limit, the lattice theory can be described
by an effective continuum Lagrangian in which the usual terms
have been supplemented by contributions proportional to powers of
the lattice spacing\cite{symanzik}. These additional pieces are
constrained by the symmetries of the lattice action, which in the
present case means that they need not respect chiral symmetry.
The enumeration of operators is identical to that carried out as
part of the improvement program\cite{luscher}, and the result is 
\begin{eqnarray}
  {\cal L}_{\rm eff} &\sim& {\cal L}_{\rm g} + 
	\bar\psi (\dirac+{m_0\over a}) \psi 
  - {\tilde m_c\over a} \bar\psi \psi
  + b_1 a \bar\psi i \sigma_{\mu\nu} F_{\mu\nu} \psi +
\nonumber\\ 
  &&\mbox{} 
  + b_2 a \bar\psi(\dirac + m)^2\psi 
  + b_3 a m \bar\psi(\dirac + m)\psi
  + b_4 a m {\cal L}_{\rm g} 
  + b_5 am^2 \bar \psi \psi + O(a^2)\,, 
\label{leff1}
\end{eqnarray}
where ${\cal L}_{\rm g}$ is the gluon Lagrangian, and $m$ is a
physical mass defined below. The symbol $\sim$ indicates that we
are not attempting to control factors of order unity. For
example, the dimensionless lattice bare mass $m_0$ appears in
${\cal L}_{\rm eff}$ multiplied by a factor \hbox{$Z_m(g^2,\ln a)
\approx 1$}, which we ignore. The dimensionless constants $\tilde
m_c$ and $b_i$ are functions of $g(a)^2$, and in general, also of
$\ln a$. We work at a fixed $a$ and thus suppress this
dependence.

The first two terms in (\ref{leff1}) are the naive continuum
limit of the lattice Lagrangian, while subsequent contributions
result from the fact that the lattice action breaks chiral
symmetry. The dominant correction at small $a$ is the additive
mass renormalization proportional to $\tilde m_c$, which is
linearly divergent as $a\to0$. We can combine the two mass terms
in the usual way by introducing the physical mass $m = (m_0 -
\tilde m_c)/a$ (again ignoring $Z$-factors with logarithmic
dependence on $a$). As we will see, $\tilde m_c$ is very close
to, but slightly different from, the critical mass $m_c(g^2)$ at
which the pion masses vanish. 

The Lagrangian can be further simplified as follows. By changing
quark variables, one can remove terms that vanish by the leading
order equations of motion\cite{sharpelat97}, which in this case
are those proportional to $b_2$ and $b_3$. Note that the
$b_4$-term
renders the effective continuum coupling constant dependent upon
the bare quark-mass, and the $b_5$-term causes the physical mass
to have a quadratic dependence on the bare quark-mass. While
these two contributions cannot be removed, they can be ignored. 
As will become clear shortly, we focus on the region in which $a
m \sim (a\Lambda)^2$, {\em i.e.} physical quark-masses of $O(a)$.
In this region, corrections proportional to $am$ are suppressed
by $a \Lambda \ll 1$ relative to the terms we keep. With these
simplifications, the effective Lagrangian becomes
\begin{equation}
  {\cal L}_{\rm eff} \sim {\cal L}_{\rm g} + 
  \bar\psi (\dirac+ m ) \psi 
  + b_1 a \bar\psi i \sigma_{\mu\nu} F_{\mu\nu} \psi + O(a^2)\,,
\label{leff2}
\end{equation}
i.e. QCD with a Pauli term. 

The next step is to write a generalization of the continuum
chiral Lagrangian that includes the effects of the Pauli term.
Without either the mass term or the Pauli term, the theory is
invariant under $SU(2)_L \times SU(2)_R$ chiral rotations, and
its low momentum dynamics is described by the chiral Lagrangian
\begin{equation}
 {\cal L}_\chi = {f_\pi^2\over 4} \,
  {\sf Tr}\, \left(\partial^\mu\Sigma^\dagger \partial_\mu\Sigma
  \right) \,.
\end{equation}
Here $\Sigma$ is an $SU(2)$ matrix-valued field that transforms
under the chiral group as
\begin{eqnarray}
  \Sigma \to L \Sigma R^\dagger \ ,
\label{chiralsym}
\end{eqnarray}
with $L$ are $R$ being independent $SU(2)$ rotations. Its vacuum
expectation value, \hbox{$\Sigma_0=\langle \Sigma\rangle$}, breaks
the chiral symmetry down to an $SU(2)$ subgroup. The fluctuations
around $\Sigma_0$ correspond to the Goldstone-bosons, 
\begin{equation}
  \Sigma = \Sigma_0\,\exp\left\{i\sum_{a=1}^3\pi_a
  \sigma_a/f_\pi\right\} \ .
\label{eq:piondef}
\end{equation}
Adding the mass term to the underlying Lagrangian explicitly
breaks chiral symmetry, and its contributions to ${\cal L}_\chi$
can be determined using a standard spurion analysis. Since we are
only interested in determining the vacuum state, we do not
consider derivative interactions. Working to second order in $m$,
the potential energy is
\begin{equation}
  {\cal V}_\chi = 
  - {c_1 \over 4 } {\sf Tr}\,\left(\Sigma + \Sigma^\dagger\right)
  + {c_2 \over 16 } \left\{
  {\sf Tr}\,\left(\Sigma + \Sigma^\dagger\right)\right\}^2 \,,
\label{vchiral}
\end{equation}
with $c_1\sim m \Lambda^3$ and $c_2\sim m^2 \Lambda^2$. We have
used the property that any term invariant under the vector
symmetry can be written as a function of 
\hbox{${\sf Tr} \left(\Sigma + \Sigma^\dagger\right)$}.

We now include the Pauli term. This is straightforward since it
transforms under chiral rotations exactly as does the mass term.
Thus its effects can be included (along with those of the mass
term) by making the substitution $m \to m + a \Lambda^2$. The
factors of $\Lambda$ are required by dimensional analysis, and 
we recall that we are dropping dimensionless coefficients of
order unity. Thus the potential is given by (\ref{vchiral}), with
coefficients
\begin{equation}
  c_1 \sim m\Lambda^3 + a \Lambda^5 \,,\qquad
  c_2 \sim m^2 \Lambda^2 + m a \Lambda^4 + a^2 \Lambda^6 \,.
\label{c12}
\end{equation}
Note that choosing $a m \sim (a \Lambda)^2$ makes the
contributions of the quark mass term and the Pauli term comparable.
This is as we would expect: if the physical mass is of $O(a)$,
then the strength of chiral symmetry breaking due to the mass
term should be comparable to that due to discretization.

Since we are keeping terms of $O(a^2)$ in $c_2$, we must also
consider corrections of $O(a^2)$ to the underlying QCD
Lagrangian. Fermion bilinears of $O(a^2)$ such as 
$a^2 \bar\psi \dirac^3 \psi$ are necessarily flavor singlets, 
and so contribute in the same way as the Pauli term, 
but suppressed by $a \Lambda$.
Four fermion operators such as $a^2\bar\psi\psi\bar\psi\psi$
can contribute directly to the $c_2$ term, and give an additional
contribution proportional to $a^2 \Lambda^6$. Thus the estimates
of (\ref{c12}) remain valid. There are of course subleading 
contributions to these coefficients, suppressed by powers of
$a\Lambda$.

In the following we determine the properties of the chiral theory, 
and in particular the pattern of symmetry breaking,
as a function of the coefficients $c_1$ and $c_2$.
Since these coefficients depend on $m_0=\tilde m_c + ma$ and on
$a$, we can map our results back onto the phase diagram in the
$m_0$--$g(a)^2$ plane. We distinguish three regions of quark
masses, each successively smaller by a factor of $a
\Lambda$:
\begin{enumerate}
\item
 Physical masses, {\em i.e.}  $m/\Lambda$ fixed and small. For
 such masses, as $a\to0$, $c_1\sim m\Lambda^3$ and $c_2/c_1\sim
 m/\Lambda$. In this case both the discretization errors and the
 contribution of the $c_2$ term can be ignored, and the symmetry
 breaking is as in the continuum.
\item
 Generic masses of $O(a)$, {\em i.e.} $a m \sim (a \Lambda)^2$.
 For these, the contribution of the $c_2$ term is suppressed by
 $a \Lambda$, but discretization errors are important in $c_1$. 
 The value of $m$ at which the extrapolated pion mass vanishes 
 is shifted by $\sim a\Lambda^2$ (corresponding to a shift of
 $O(a^2\Lambda^2)$ in $m_0$). It is useful to introduce a shifted
 mass, $m'= m - a \Lambda^2$, (where, we stress, all constants
 have been set to unity), defined so that $c_1$ vanishes when
 $m'=0$. In terms of this new mass the coefficients are
\begin{equation}
  c_1 \sim m'\Lambda^3\,,\qquad
  c_2 \sim m'^2 \Lambda^2 + m' a \Lambda^4 + a^2 \Lambda^6 \,.
\label{c12new}
\end{equation}
\item
 Shifted masses of $O(a^2)$, {\em i.e.} $a m' \sim (a
 \Lambda)^3$. For such masses, the coefficients can be 
 simplified to
\begin{equation}
  c_1 \sim m'\Lambda^3\,,\qquad
  c_2 \sim a^2 \Lambda^6 \,.
\label{c12newest}
\end{equation}
 The crucial point is that $c_1\sim c_2$, and so the two terms 
 in the potential are comparable. As explained below, competition
 between the two terms can lead to Aoki phases.
\end{enumerate}
It is immediately apparent from this discussion that the
width of the region in which new phenomena can occur, due
to competition between the two terms in the potential, is 
\hbox{$\Delta m_0 \sim a \Delta m' \sim (a \Lambda)^3$}.
This is the result announced in the introduction.

One might be concerned that higher order contributions to the
potential will become important if the
first two terms are comparable to one another.
A simple analysis shows that the coefficients of the next order
terms  are generically of size $c_3 \sim a^3 \Lambda^7$. 
Compared to the first two terms, these are suppressed by $a
\Lambda$, and can be ignored, except for the
possibility of a small region of width $\Delta m_0 \sim a^4$ in
which all three terms could cancel.
Similarly, an $a^2 \Lambda^2$ contribution to $c_1$ will shift
$m'$ by $O(a^2)$ but will not otherwise change the analysis.

To determine the condensate $\Sigma_0$ we must minimize the
potential energy (\ref{vchiral}). Writing \hbox{$\Sigma = A + i
{\bf B} \cdot{\lpmb \sigma}$} with \hbox{$A^2 + {\bf B}^2 = 1$}, 
the potential becomes
\begin{eqnarray}
  {\cal V}_\chi&=& -c_1 A + c_2 A^2  \ ,
\label{lmint}
\end{eqnarray}
where the parameter $A$ is constrained to lie between $-1$ and $+1$
inclusive. Note that the $SU(2)_V$ group action, in which $L=R$
in (\ref{chiralsym}), leaves $A$ invariant and rotates ${\bf B}$
by an orthogonal transformation. Hence, when the vacuum state
\hbox{$\Sigma_0 =A_0 + i {\bf B_0} \cdot{\lpmb \sigma}$} develops
a non-zero ${\bf B}_0$, the flavor symmetry breaks spontaneously
to a $U(1)$ subgroup defined by $\exp\left\{i\theta\, \hat{\bf
B}_0 \cdot {\lpmb \sigma} \right\}$, with $\hat{\bf B}_0$ being
the unit vector in the direction of ${\bf B}_0$.  A non-zero
value of ${\bf B}_0$ can occur only when $|A_0|$ is strictly less
than one. 

\begin{figure}
\centerline{
\epsfxsize=85mm
\epsfbox{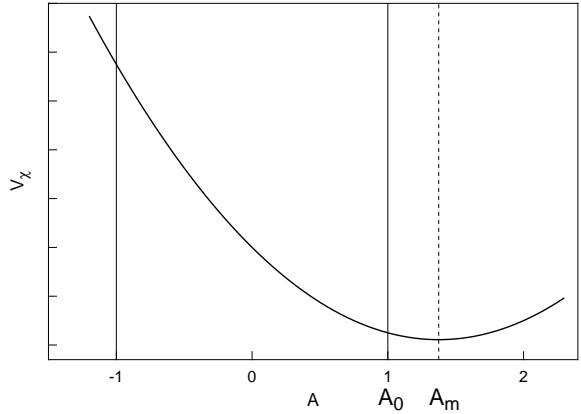}
}
\caption{The potential ${\cal V}_\chi$ vs. $A$ for $c_2>0$ and
$|A_m|>1$. The vacuum is at $A_0=1$. 
}
\label{fig:parab1}
\end{figure}

\begin{figure}
\centerline{
\epsfxsize=85mm
\epsfbox{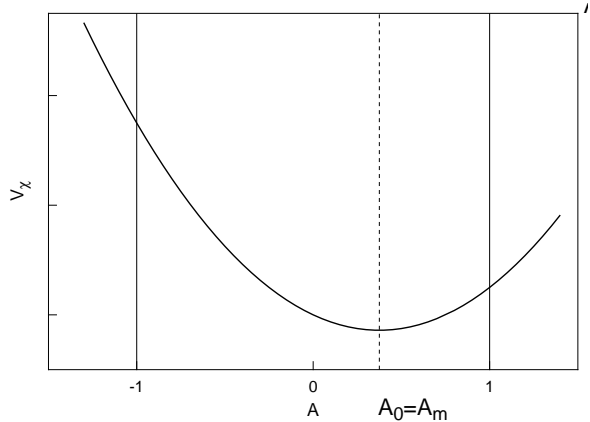}
}
\caption{The potential ${\cal V}_\chi$ vs. $A$ for $c_2>0$ and
$|A_m|<1$. The vacuum is at $A_0=A_m$. 
}
\label{fig:parab2}
\end{figure}

Based on the discussion above, we treat $c_2$ as a constant of
$O(a^2)$, while $c_1$ varies linearly with the bare quark-mass.
We do not know the sign of $c_2$. It turns out that if $c_2$ is
positive, we reproduce the properties of the Aoki phase, and so
we discuss this case first. The potential is then a parabola with
its minimum at $A_m=\epsilon$, where we have defined the
parameter $\epsilon=c_1/2 c_2 \sim m'/(a^2 \Lambda^3)$. If this
minimum lies outside the range $-1$ to $1$, then $A_0$ is forced
to one of the boundary points $|A_0|=1$. This corresponds to
$\Sigma_0 = \pm 1$, and hence the vector symmetry $SU(2)_V$ is
not spontaneously broken. This situation is illustrated in 
Fig.~\ref{fig:parab1}. If instead the minimum satisfies
$|A_m|<1$, then the vacuum is determined by $A_0=A_m$ and ${\bf
B}_0 \ne 0$, as illustrated in Fig.~\ref{fig:parab2}. Since ${\bf
B}_0 \ne 0$, flavor symmetry is spontaneously broken to $U(1)$,
and the region \hbox{$-1 < \epsilon < 1$} thus has the
properties of the Aoki phase.


\begin{figure}
\centerline{
\epsfxsize=85mm
\epsfbox{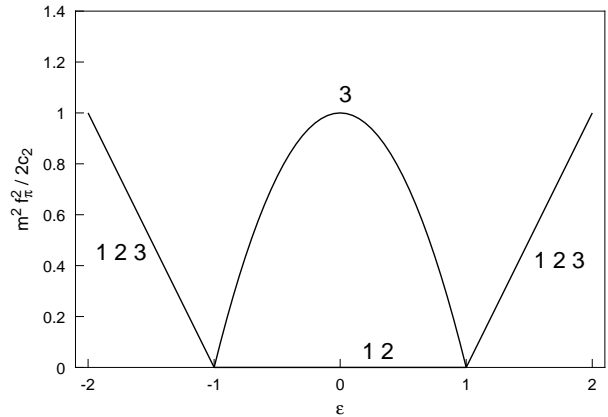}
}
\caption{%
Pion masses as a function of $\epsilon$ for $c_2>0$. The curves
are labeled by the flavor of the corresponding pion. 
}
\label{fig:masseps}
\end{figure}

Let us examine the spontaneous flavor breaking in more detail. In
the Aoki phase, the direction of ${\bf B}_0$ is determined by the
source term. To make contact with the numerical simulations, we
assume a source that aligns ${\bf B}_0$ in the 3-direction, and
adopt the new parameterization $\Sigma_0 = \cos\theta_0 + i
\sin\theta_0 \sigma_3$. The angle $\theta_0$ is determined by
$\epsilon$, as explained above,

\begin{equation}
\cos\theta_0 = \left\{ 
  \begin{array}{cr} 
  -1 & ~~~~\epsilon\le -1 \\
  \epsilon & ~~~~ -1 \le \epsilon \le 1 \\
  +1 & ~~~~ 1\le \epsilon
  \end{array} \right. \ .
\label{costzero}
\end{equation}
The vacuum smoothly interpolates between the two flavor-symmetric
values. To determine the pion masses, we expand $\Sigma$ about
the condensate (\ref{eq:piondef}), finding
\begin{equation}
  A = \cos\theta_0 - \frac{\sin\theta_0}{f_\pi}\, \pi_3
  - \frac{\cos\theta_0}{2f_\pi^2}\, \sum_{a=1}^{3} \pi_a^2 +
  {\cal O}(\pi_a^3) \ .
\label{Aeqa}
\end{equation}
The potential becomes
\begin{eqnarray}
  {\cal V}_\chi = \left\{ \begin{array}{cr}
  \frac{\textstyle c_2}{\textstyle f_\pi^2}\, (1 - \epsilon^2)\,
  \pi_3^2  - c_2 \epsilon^2 + {\cal O}(\pi_a^3) 
  & ~~~~~~~|\epsilon| < 1 \\
  \frac{\textstyle c_2}{\textstyle f_\pi^2}\,
  \left(|\epsilon|-1\right)
  \sum_a \pi_a^2 & ~~~~~~~ 1 \le |\epsilon| 
  \end{array} \right. \ ,
\end{eqnarray}
and hence the pion masses are
\begin{mathletters}
\begin{eqnarray}
  m_1^2=m_2^2 = 0\,,\quad  
  \frac{m_3^2 f_\pi^2}{2 c_2} &=& 1-\epsilon^2 
  ~~~~~\, {\rm for}~~~ |\epsilon| \le 1
\\ 
  \frac{m_a^2 f_\pi^2}{2 c_2} &=& |\epsilon|-1
  ~~~~~ {\rm for}~~~ |\epsilon| \ge 1  \ .
\end{eqnarray}
\label{eq:mpiresults}
\end{mathletters}
The results are shown in Fig.~\ref{fig:masseps}. We see that the
pions $\pi_{1,2}$ are the Goldstone  bosons of the broken flavor
symmetry within the Aoki phase, and that all three pions are
massless on the phase boundaries.

In summary, we have reproduced the phenomenology of the Aoki
phases using the chiral Lagrangian. In addition, our calculation
makes three predictions:
\begin{enumerate}
\item
 As already mentioned above, for small enough $a$, the width of
 the Aoki phase should scale as $\Delta m_0 \sim a^3$. This
 should hold up to logarithmic corrections, which we have ignored
 throughout. 
\item
 From (\ref{eq:mpiresults}), the mass of the $\pi_3$ meson within
 the Aoki phase is predicted in terms of the form outside the
 phase. For example, the slope of $m_3^2$ versus the
 dimensionless bare mass $m_0$ should be
 a factor of two larger as the phase boundary is approached from
 within compared to approaching it from without. In both cases,
 $m_3 \sim a$ when $\epsilon \sim {\cal O}(1)$.
\item
 The form of the spectral density of the overlap Hamiltonian at 
 zero eigenvalue can be determined. To do this we note that
\begin{equation}
  \langle\bar\psi_a i\gamma_5 \psi_b \rangle \propto 
  i\left[ \Sigma_0 - \Sigma_0^\dagger \right]_{ab} \,.
\label{latticecondeq}
\end{equation}
On the lattice, the l.h.s. is proportional to $\rho(0;m_0)$,
while we have found that the r.h.s. is proportional to
$\sin\theta_0$, and therefore
\begin{equation}
  \rho(0;m_0) \propto \sqrt{1-\epsilon^2}
  \propto m_3 \ .
\end{equation}
\end{enumerate}
It should be possible to test some of these predictions.

\begin{figure}
\centerline{
\epsfxsize=85mm
\epsfbox{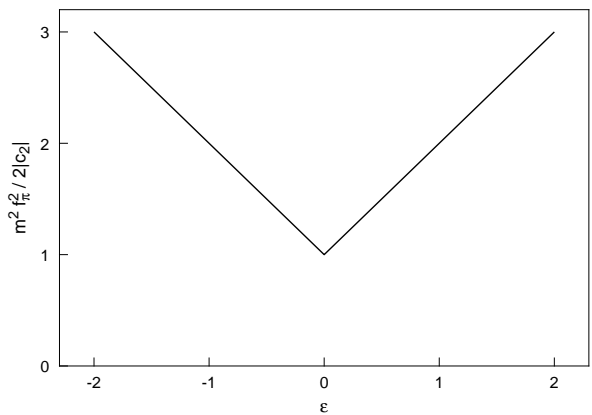}
}
\caption{%
Pion masses as a function of $\epsilon$ for $c_2<0.$
}
\label{fig:massnobrk}
\end{figure}

\bigskip\bigskip
We now return to the possibility that the coefficient $c_2$ is
negative. The potential is then an inverted parabola, and the
extremum $A_m$ is an absolute maximum. This means that the vacuum
state is always at the edge of the allowed range of $A$, with
$\Sigma_0=+1$ for $c_1> 0$ and $\Sigma_0=-1$ for $c_1<0$.
The potential can be written
\begin{equation}
  {\cal V}_\chi =
  \frac{|c_2|}{f^2_\pi}\left(1 + |\epsilon|\right)
  \,\sum_{a=1}^3 \pi^2_a \,+\, 
  c_2 - |c_1| + {\cal O}(\pi_a^3)\ .
\end{equation}
Thus the flavor symmetry is not spontaneously broken for any value
of $\epsilon$, and all three pions have the same non-zero mass,
\begin{equation}
  \frac{m_a^2 f^2_\pi}{2|c_2|} = 1 + |\epsilon| \ .
\end{equation}
This situation is illustrated in Fig.~\ref{fig:massnobrk}.
Note that $m_a\sim a$ for $\epsilon \sim {\cal O}(1)$.

The previous analysis cannot choose between the $c_2>0$ and
$c_2<0$ cases --- for this we must rely on simulations. As noted
in Sec.~\ref{sec:cons}, all numerical observations to date are 
consistent with there being spontaneous flavor and parity
breaking at non-zero lattice spacing, and hence with $c_2$ being
positive. Thus it appears that an Aoki phase with a width of
order $a^3$ will persist until the continuum is reached. There
are, however, two caveats that we should mention. First, removing the
quenched (or partially quenched) approximation
could, in principle, lead to a change in the sign of $c_2$. 
Second, it is logically possible that, for the
lattice spacings studied to date, terms of higher order in $a$
are important and conspire with the terms we considered above in
such a way that the effective value of $c_2$ changes sign as the
continuum limit is approached. In this case the Aoki phase would
disappear for small enough lattice spacing. 

It is interesting to consider in what way the previous analysis
changes when one uses improved Wilson fermions. The
surprising answer is that the analysis is essentially unaltered,
even if one removes all errors proportional to $a$ using
non-perturbative improvement conditions. The reason is that the
key discretization error is the $O(a^2)$ contribution to $c_2$
[see (\ref{c12})], and this remains after improvement. Of course,
the sign of the coefficient $c_2$ can change, so we cannot
predict whether there will be an Aoki phase. What we can say
is that one or other of
the possibilities discussed above will apply.\footnote{In fact the numerical
evidence of Ref.~\cite{EHNII} suggests an Aoki phase also for
non-perturbatively improved fermions.}
This is surprising because in both scenarios the non-vanishing pion
masses are proportional to $a$ for $\epsilon\sim O(1)$, while one
would have naively expected the masses to be proportional to
$a^2$. We see nothing wrong with this result, however. In
particular, for any fixed physical quark-mass, the discretization
errors are of $O(a^2)$ for small enough lattice spacing.

\section{Conclusions}
\label{sec:conc}

We have considered the issue of spontaneous flavor and parity
symmetry breaking in lattice QCD, and its relation to chiral
symmetry breaking in the continuum. Contrary to the claims of
Bitar {\em et al.} \cite{BHN}, we have shown that flavor-parity
breaking can occur at any non-zero lattice spacing. Indeed, our
chiral Lagrangian analysis shows that such symmetry breaking is
one of only two options available to the theory as the continuum
limit is approached. If this option is chosen, as is indicated by
present numerical evidence, then our calculation shows how the
breaking of flavor and parity at non-zero lattice spacing goes
over smoothly into chiral symmetry breaking in the continuum
limit. This is the sense in which we agree with the proposal of
Bitar {\em et al.} that one can think of flavor-parity breaking
on the lattice as corresponding to chiral symmetry breaking in
the continuum.

Based on this proposal, Bitar {\em et al.} suggest obtaining
the continuum chiral condensate, 
\hbox{$C_0 = \langle\bar u u\rangle + \langle\bar d d\rangle$},
in the following way. Calculate the condensate 
\hbox{$\langle \bar\psi i\gamma_5 \sigma_3 \psi \rangle$}
in the infinite volume 
limit at non-zero lattice spacing; average the result over the
Aoki-phase; and then take the continuum limit. Their major
point, with which we agree, is that the lattice condensate
is a good order parameter, and in particular is free from
additive renormalization. It follows from our analysis, however, that
one should not average over the Aoki phase, but rather take the
maximum value of the lattice condensate within the Aoki phase,
and extrapolate this to the continuum limit. 
If one were to average over the Aoki phase then one would find
a value for the continuum condensate too small by a factor of $\pi/4$.

To understand this, recall that the condensate takes the
form $\Sigma_0 = \cos\theta_0 + i \sin\theta_0 \sigma_3$
in the Aoki phase, and so
\begin{eqnarray}
\langle\bar\psi \psi\rangle &=& 
C_0 {\sf Tr}\left({\Sigma_0+\Sigma_0^\dagger\over 4}\right)
= C_0 \cos\theta_0 \,,\\
\langle\bar\psi i \gamma_5 \sigma_3 \psi\rangle &=& 
C_0 {\sf Tr}\left({\Sigma_0-\Sigma_0^\dagger\over 4i}\right)\sigma_3
= C_0 \sin\theta_0 \,.
\end{eqnarray}
Thus at a general position in the Aoki phase, one can obtain
continuum condensate from
\begin{equation}
C_0^2 = \langle\bar\psi \psi\rangle^2 +
\langle\bar\psi i \gamma_5 \sigma_3 \psi\rangle^2 \,.
\end{equation}
On the lattice, however, one can only measure the second of these
condensates, and this equals $C_0$ only when
$\epsilon=\cos\theta_0=0$, {\em i.e.} when the lattice condensate
takes its maximum value. In other words, the lattice condensate
picks out a particular component of the chiral condensate, and
only when the former is maximal are the two condensates aligned.

One of the predictions of our analysis is that the width of the 
symmetry-broken phase grows rapidly as the lattice spacing is
increased, $\Delta m_0 \sim a^3$. According to Aoki's picture,
there is a critical lattice spacing at which the five ``fingers''
of the Aoki phase merge into a single phase of width $\Delta m_0
\sim 6$  (see Fig.~\ref{fig:aoki}). It is not clear from
numerical evidence at what coupling this occurs. At $g^2=1$, in
the quenched approximation, Ref.~\cite{aku} finds evidence
for all five fingers, while Ref.~\cite{EHNII} finds only a single
phase down to $m_0=-2$. We stress, however, that either
possibility is consistent with our results. Our prediction is
that there is some critical coupling below which the Aoki phase
$B$ splits into finger-like regions of decreasing width,
surrounded by islands of unbroken phase $A$, as in
Fig.~\ref{fig:aoki}. Once one has entered this region then our
other predictions, for the pion masses and the condensate, apply
as well.

As explained in Ref.~\cite{EHNII}, practical applications of
domain-wall fermions probably require that the Aoki phase has
a small width. Our results strongly suggest that this will be the case for
small enough lattice spacing.

\section*{Acknowledgments}

We thank Khalil Bhitar, Rajamani Narayanan, Urs Heller and Larry Yaffe
for helpful comments and discussions. This research was supported
in part by U.S. Department of Energy grant DE-FG03-96ER40956. 

\newpage
\begin{appendix}

\section{The Inapplicability of the Vafa-Witten theorems}
\label{app:vw}

The condensate (\ref{condlat}) breaks both flavor and parity,
in apparent contradiction with Refs.~\cite{vw1} and~\cite{vw2}, 
respectively. In this appendix we give a brief explanation of 
why these references do not, in fact, rule out such a condensate.

The argument of Ref.~\cite{vw1} against flavor-breaking in the
continuum does not apply for massless fermions because the
spectrum of the Dirac operator has a non-vanishing density of
zero eigenvalues. As has been discussed in Ref.~\cite{BHN}, 
the relevant operator on the lattice is the overlap
Hamiltonian, $\gamma_5 W(-m_0)$, and the Vafa-Witten argument
also does not apply if this has a non-vanishing density of zero
eigenvalues, which it does in the Aoki phase. These failures
are demonstrated by the explicit calculations of
Appendix~\ref{app:calcs}.

The loophole in the argument against spontaneous parity-breaking
is simply that the analysis says nothing about fermionic bilinears.
The bulk of Ref.~\cite{vw2} is devoted to proving that parity-odd
bosonic operators do not receive vacuum expectation values. In
Footnote~7, however, Vafa and Witten claim that the theorem can
be extended to include fermionic bilinears, since, upon
integrating out the fermions, the fermionic operators can be
replaced by bosonic operators, and for these their original
argumentation applies. The example they consider is the
flavor-singlet operator $\bar q i
\gamma_5 q$, which is replaced by \hbox{$X={\sf Tr} i\gamma_5
S_A(x;x)$}, with $S_A(x;y)$ being the fermionic propagator in the
background $A$. This extension of their argument is not valid,
however. The operator $X$ is either ill-defined if there
are fermionic zero modes, or it simply vanishes because the
eigenvalues of $\gamma_5 S_A(x;y)$ come in opposite-sign pairs.

Properly speaking, one must calculate the vacuum expectation
value by adding a source, and only take the source strength to
zero at the end. The presence of the source regulates the
corresponding effective bosonic operator $X$. However, for
two flavors of quarks, it also invalidates an essential assumption
of Ref.~\cite{vw2}, namely that $X$ be imaginary in Euclidean
space. Indeed, one can explicitly calculate the form of $X$ using
a spectral decomposition of the corresponding fermion operator.
For the condensate of interest, (\ref{condlat}), for which one
adds the source (\ref{source1}), this calculation is done in
Appendix~\ref{app:calcs}. We find that the corresponding gluonic
operator $X$ is purely real, and so the Vafa-Witten argument does
not apply.\footnote{
For the flavor-singlet parity-odd operator
$\bar q i \gamma_5 q$, on the other hand, the effective bosonic
operator $X$ in the presence of the source ${\cal L}_{\rm source}
= h \bar q \exp\left[i \theta \gamma_5\right] q$ is not pure
real, and the loophole in applying Ref.~\cite{vw2}
is more subtle. Note that the source can be transformed into the
mass term $h\bar q q$ by an axial $U(1)$ transformation, which,
because of the axial anomaly, introduces a $\theta \tilde F F$ term. 
The source therefore does not have the usual effect of picking
a particular direction for the condensate, but instead 
selects a different theory characterized by a
parity-violating \hbox{$\theta$-vacuum}.
This is true even for infinitesimal source strength,
and thus Ref.~\cite{vw2} does not imply
that the fermionic parity-breaking condensate $\langle \bar q i
\gamma_5 q\rangle$ vanishes.
}
This point has also been noted in Ref.~\cite{aokigocksch}.

It should be noted, however, that Vafa and Witten's argument does
still imply that all parity-odd operators composed only of gluon
fields do not have vacuum expectation values. This suggests that
there exists a modification of parity which acts in the same way
as parity on the gluon fields, but differently on fermions, and
which is unbroken by the condensate in the Aoki phase. Such a
transformation does exist: it is the product of parity with a
discrete flavor transformation,
\begin{equation}
P^\prime = P \otimes z \,,\qquad z=i \sigma_1 \in SU(2) \,.
\end{equation}
The flavor transformation has no effect on the gluon fields.
Note that $(P^\prime)^2 = -1$ when acting on fermion fields, so
that $P^\prime$ generates a $Z_4$ group. When
acting on the bilinear $\bar\psi i\gamma_5 \sigma_3 \psi$,
however, its square is the identity. Note also that other choices
of $z$ are possible, but all are related by transformations in
the unbroken $U(1)$ flavor subgroup.

\section{Calculation of the Condensates}
\label{app:calcs}

In this appendix we compute the condensates $\langle
\bar\psi_b\Gamma \psi_a \rangle$ in both the continuum
and on the infinite-volume lattice with Wilson fermions. We work
in two-flavor QCD with a common bare mass, and we denote the
quark field with flavor index $a$ by $\psi_a$. The quantity
$\Gamma$ is an appropriately chosen spinor-matrix, which in the
continuum can be either the unit matrix ${\bf 1}$ or $i\gamma_5$,
while on the lattice we consider only $\Gamma=i\gamma_5$. As
discussed in Sec.~\ref{sec:cons}, the condensates are defined by
adding to the Lagrangian the source term
\begin{equation}
  {\cal L}_{\rm source}({\lpmb \theta}, h) =
  h \bar\psi \exp\left[i {\lpmb \theta}
  \cdot {\lpmb \sigma}\gamma_5\right] \psi  \ ,
\label{apsource2}
\end{equation}
computing $\langle \bar\psi_b\Gamma \psi_a \rangle_h$ for
non-zero $h$, and then taking the $h\to 0^+$ limit. We will
let $\theta$ denote the modulus of ${\lpmb \theta}$, while
$\hat{\lpmb \theta}$ will represent a unit vector in the
direction of ${\lpmb \theta}$. Without loss of generality we 
can take $h > 0$ and $0\le \theta < \pi$ (so that $\sin\theta 
\ge 0$).

\subsection{The continuum theory}

We first consider the massless continuum theory with the fermionic
Lagrangian ${\cal L}_f = \bar\psi_a \dirac \psi_a$. It will be
convenient to place the system in a finite box of four-volume
$V_4$, so that the Euclidean (anti-Hermitian) Dirac-operator
$\dirac$ has discrete eigenvalues $\pm i \lambda_n$ with
$\lambda_n\ge 0$, and at the end of the calculation the 
infinite-volume limit followed by the $h\to 0^+$ limit will be
taken. For simplicity, let us first work out the flavor diagonal
case $\hat{\lpmb \theta}= \hat{\lpmb z}$ with arbitrary $\theta$.
After performing the fermionic path integrals, the condensates in
the presence of the source can be written 
\begin{equation}
  \langle \bar\psi_b \Gamma \psi_a \rangle_h =
  \int {\cal D}A\, \exp\left\{-S_g[A]\right\}\,
  \left[{\rm det}\,(\dirac + h)\right]^2 \, 
  \langle \bar\psi_b \Gamma \psi_a \rangle_{A,h} \ ,
\label{fullcond}
\end{equation}
with $S_g[A]$ being the pure gluonic Euclidean action, and 
where the fixed-background condensates take the form
\begin{eqnarray}
  \langle \bar\psi_b \Gamma \psi_a \rangle_{A,h} =
  \frac{1}{V_4}\int d^4 x \,\langle \bar\psi_b \Gamma 
  \psi_a(x) \rangle_{A,h}  = -\frac{1}{V_4}\,{\sf Tr} 
  \frac{\Gamma\,\delta_{ab}}{\dirac + h \exp\left\{i 
  \theta \gamma_5\sigma_{3\, ab}\right\}} \ .
\label{acond}
\end{eqnarray}
The trace includes a sum over color, spin, and
space-time, and it can be written
\begin{eqnarray}
  {\sf Tr}\,\frac{\Gamma}{\dirac + h  \exp\left\{ 
  \pm i \theta \gamma_5\right\}}  = \sum_{\lambda_n\ge 0} 
  \frac{2 h}
  {\lambda_n^2 + h^2}\, {\scriptscriptstyle \times}
  \left\{\begin{array}{rl} 
     \cos\theta & ~~{\rm for}~~\Gamma={\bf 1}  \\
  \pm\sin\theta & ~~{\rm for}~~\Gamma=i\gamma_5 \\
  \end{array}\right. \ ,
\label{trcond}
\end{eqnarray}
where, in computing the fermionic determinant in 
(\ref{fullcond}) and the trace (\ref{trcond}),
we have used the fact that 
\begin{mathletters}
\label{diracev}
\begin{eqnarray}
  \dirac + m_1 + i \mu_1 \gamma_5 
   ~~~&&{\rm has~eigenvalues}~~
  \Lambda_n^\pm= m_1 \pm i \sqrt{\lambda_n^2 + \mu_1^2}
\label{diraceva} \\
  \gamma_5(\dirac + m_1 + i \mu_1 \gamma_5)
   ~~~&&{\rm has~eigenvalues}~~
  \Lambda_n^\pm= i \mu_1 \pm \sqrt{\lambda_n^2 + m_1^2} \ .
\label{diracevb}
\end{eqnarray}
\end{mathletters}%
Since we are working with two-flavor QCD, the fermionic
determinant is real, positive, and $\theta$-independent. 

In infinite volume, the eigenvalues of the Hermitian operator
$-i\dirac$ become continuous and are described by a spectral
density per unit four-volume $\tilde \rho_A(\lambda)$.
Since $-i\dirac$ and $\gamma_5$ anti-commute, the eigenstates
of $-i\dirac$ come in pairs $\phi_\lambda(x)$ and
$\gamma_5\phi_\lambda(x)$ with eigenvalues $\lambda$ and 
$-\lambda$ respectively. Therefore, $\tilde\rho_A(\lambda)$ is
symmetric in $\lambda$ and we can write
\begin{eqnarray}
  \lim_{h\to 0^+}\lim_{V_4\to\infty} {\frac{1}{V_4}\sum_n 
  \frac{2 h}{\lambda_n^2 + h^2}} = \lim_{h\to 0^+}
  \int_{-\infty}^\infty d\lambda\, \tilde\rho_A(\lambda) \,
  \frac{h}{\lambda^2+h^2} = \pi \tilde\rho_A(0) \ ,
\label{ratfun}
\end{eqnarray}
where we have used the relation
\begin{eqnarray}
  \lim_{h\to 0^+} \frac{1}{\lambda \pm i h} = 
  \mp i \delta(\lambda) + P\frac{1}{\lambda} \ , 
\label{delpp}
\end{eqnarray}
with $P\frac{1}{\lambda}$ denoting the principal part under 
the $\lambda$ integral. The principal-part contribution vanishes,
however, because of the symmetry in $\lambda$. In the flavor
diagonal case we thus find
\begin{eqnarray}
  \lim_{h\to0^+} \langle \bar\psi_b \Gamma 
  \psi_a \rangle_{A,h} = -\pi\, \tilde\rho_A(0)\,  
  {\scriptscriptstyle \times} 
  \left\{\begin{array}{ll} 
  \cos\theta \,\delta_{ab}& ~~{\rm for}~~\Gamma={\bf 1}  \\
  \sin\theta \,\sigma_{3\,ab}& ~~{\rm for}~~\Gamma=i\gamma_5 \\
  \end{array}\right. \ .
\label{fdiag}
\end{eqnarray}

The condensate in the presence of the source
(\ref{apsource2}) with an arbitrary flavor orientation
$\hat{\lpmb \theta}$ can be obtained from (\ref{fdiag}) by 
an appropriation spinor-rotation. If we write
$\hat{\lpmb \theta}=(\sin\alpha \cos\beta,\sin\alpha \sin\beta,
\cos\alpha)$ with \hbox{$\hat{\lpmb \Theta}=(\sin(\alpha/2)
\cos\beta,\sin(\alpha/2) \sin\beta, \cos(\alpha/2))$},
then $R \equiv i\hat{\lpmb \Theta}\cdot{\lpmb \sigma}$ satisfies
\begin{eqnarray}
  R \sigma_3 R^\dagger = \hat{\lpmb \theta}\cdot{\lpmb 
  \sigma} \ ,
\label{sig3xform}
\end{eqnarray}
and we see that $R$ is a \hbox{spin-1/2} representation of the
\hbox{3-dimensional} rotation from $\hat{\lpmb z}$ to $\hat{\lpmb
\theta}$. As $\gamma_5$ does not act in the flavor space and
since its square is unity, the matrix $R$ also transforms the
diagonal source with $\hat{\lpmb \theta}=\hat{\lpmb z}$ to the
general source (\ref{apsource2}) using the same transformation as
(\ref{sig3xform}), {\em i.e.}
\begin{eqnarray}
  R \exp\left[i \theta \sigma_3 \gamma_5\right]R^\dagger 
  = \exp\left[i {\lpmb \theta}\cdot{\lpmb \sigma}
 \gamma_5\right] \ .
\label{sourcexform}
\end{eqnarray}
Using (\ref{sourcexform}), the diagonal condensate
(\ref{fdiag}) can be rotated to a general orientation
$\hat{\lpmb \theta}$ to produce
\begin{mathletters}
\label{nonfdiag}
\begin{eqnarray}
  \langle \bar\psi_b \psi_a 
  \rangle_A &=& -\pi\, \tilde\rho_A(0)
  \cos\theta\,\delta_{ab}
\label{nonfdiaga}\\
  \langle \bar\psi_b i\gamma_5 \psi_a \rangle_A
  &=& -\pi\, \tilde\rho_A(0)\sin\theta \,
  [\hat{\lpmb \theta}   \cdot {\lpmb \sigma}]_{ab} \ .
\label{nonfdiagb}
\end{eqnarray}
\end{mathletters}
The condensates $\langle \bar\psi_b \Gamma \psi_a \rangle$ can
now be obtained with the help of (\ref{fullcond}), and they
take the same form as (\ref{nonfdiag}), except that one
uses the gauge averaged spectral density
\begin{equation}
  \tilde\rho(\lambda) = 
  \int {\cal D}A\, \exp\left\{-S_g[A]\right\}\,
  \left[{\rm det}\,\dirac\right]^2 \, 
 \tilde\rho_A(\lambda) \ .
\end{equation}
Equation (\ref{condcont}) now follows. 

It is also interesting to consider the condensates for a non-zero
common mass $m$ with fermionic Lagrangian \hbox{${\cal L}_f
=\bar\psi_a \left(\dirac+m\right)\psi_a$}. As with the massless
case in (\ref{ratfun}), a condensate before the $h\to 0^+$ limit
has been taken can be expressed as a spectral integral of some
rational function. The numerators of the scalar and
pseudo-scalar condensates are proportional to $m + h\cos\theta$
and $h\sin\theta$ respectively, while the denominators of
both condensates are \hbox{$\lambda^2+h^2+m^2+ 2mh\cos\theta$}.
The non-zero mass $m$ renders the $h\to0^+$ limit safe for all
eigenvalues $\lambda$, thereby producing the $\theta$-independent
expressions
\begin{mathletters}
\label{condm}
\begin{eqnarray}
  \langle \bar\psi_b \psi_a 
  \rangle_A &=& -\int_{-\infty}^\infty d\lambda\,
  \tilde\rho_A(\lambda)\,\frac{m}{\lambda^2 + m^2}\,
  \delta_{ab}
\label{condma} \\
  \langle \bar\psi_b i\gamma_5 \psi_a \rangle_A &=& 0 \ .
\label{condmb}
\end{eqnarray}
\end{mathletters}
Of course, taking the $m\to0$ limit first would give
(\ref{nonfdiag}), and we see that the $m\to 0$ and $h\to0^+$
limits do not commute. Note, however, that (\ref{condma}) is
consistent with the fact that flavor is not spontaneously
broken for massive quarks\cite{vw1}, since it implies that
$\langle\bar\psi \sigma_3 \psi\rangle=0$, while (\ref{condmb})
is consistent with there being no spontaneous flavor or parity breaking.

\subsection{The infinite-volume lattice theory}

While much of the corresponding calculation on the infinite-volume 
lattice with Wilson fermions is similar to the continuum case,
there are some key distinctions which we outline in this section.
For one thing, the Wilson-Dirac operator $W(m_0)$ has complex
eigenvalues and therefore does not possess a spectral density.
This suggests that one tries to relate the condensates to the
overlap Hamiltonian $H(-m_0)=\gamma_5W(m_0)$, which is Hermitian
and therefore has eigenvalues constrained to lie along the real
axis. For a fixed background configuration $U$, we will write the
discrete finite-volume eigenvalues of $\gamma_5 W(m_0)$ as
$\lambda_n(m_0;U)$, while in infinite volume the corresponding
spectral density will be denoted by $\rho_U(\lambda;m_0)$.
Therefore, the fermionic Lagrangian for the quarks $\psi_a$ with
a common bare mass $m_0$ can be written ${\cal L}_f =\bar\psi_a\,
W(m_0) \psi_a = \psi'_a H(-m_0) \psi_a$, where we now consider
$\psi_a$ and $\psi'_a=\bar\psi_a\gamma_5$ as the independent
Grassmann variables. While the Wilson term explicitly breaks 
the chiral symmetry, there is nonetheless an exact $SU(2)$ 
flavor symmetry. We should point out that, on the 
lattice, we only calculate $\langle\bar\psi_b \Gamma \psi_a 
\rangle$ with $\Gamma=i\gamma_5$. Unlike the continuum, the
condensate with $\Gamma=1$ is not proportional to the spectral
density, since the eigenvectors of $\gamma_5 W(m_0)$ with
opposite-sign eigenvalues are not connected by $\gamma_5$.

We first consider the flavor diagonal case $\hat{\lpmb \theta}=
\hat{\lpmb z}$ with arbitrary $\theta$. As in the continuum, we
place the system in a finite box of four-volume $V_4$, calculate
the condensate with
source (\ref{apsource2}), and then take the infinite-volume
limit followed by the  $h\to0^+$ limit. The condensate $\langle
\bar\psi_b i \gamma_5 \psi_a\rangle_h$ vanishes when $\theta=0$,
and we therefore concentrate on $\theta > 0$. After
performing the fermionic path integrals over $\psi$
and $\psi'=\bar\psi\gamma_5$, we find
\begin{equation}
  \langle \bar\psi_b i\gamma_5 \psi_a \rangle_h =
  \int {\cal D}U \exp\left\{-S_g[U]\,\right\}\,
  \,\bigg|{\rm det}\left[\gamma_5W(m_0) + h \gamma_5 \exp
  \left\{i\theta\gamma_5\right\}\right]\bigg|^2\,
  \langle \bar\psi_b i\gamma_5 \psi_a \rangle_{U,h} \ ,
\label{latfullcond}
\end{equation}
with $S_g[U]$ being the Euclidean Wilson action, and 
where the background condensates take the form
\begin{eqnarray}
  \langle \bar\psi_b i\gamma_5 \psi_a\rangle_{U,h} =
  -\frac{1}{V_4}\,{\sf Tr}\, \frac{i\delta_{ab}}
  {\gamma_5W(m_0) + h \gamma_5 \exp
  \left\{i\theta\gamma_5\sigma_{3\, ab}\right\}} \ ,
\label{Ucondtr}
\end{eqnarray}
with the trace including a sum over spin, color, and space-time.
Note that in two-flavor QCD, the fermionic determinant factors
are real and positive, although unlike in the continuum there is
explicit $\theta$-dependence at non-zero $h$. Using the chiral
basis for the gamma-matrices, in which $\gamma_5={\rm
diag}\,(1,-1)$, the overlap Hamiltonian takes the general
form\cite{ol}
\begin{eqnarray}
  \ham(-m_0) = \gamma_5 W(m_0) = \pmatrix{\bmat + m_0 & \cmat \cr
  \cmat^\dagger & - \bmat - m_0} \ ,
\label{hamil}
\end{eqnarray}
where the two-component spinor operator $C$ is a discretized 
chiral Dirac operator, and $B$ arises from the Wilson term.
{}From this expression, the operator appearing in the denominator
of (\ref{Ucondtr}) can be written
\begin{eqnarray}
  \begin{array}{rr}
     \gamma_5W(m_0) + h \gamma_5 \exp\left\{i\theta\gamma_5
  \sigma_{3\, ab}\right\} = \gamma_5 W(M) + i h\sin\theta\,
  \sigma_{3 ab}\, \\
   ~~~{\rm with}~~ M=m_0+h\cos\theta \ , \\
  \end{array} 
\end{eqnarray}
and it therefore has eigenvalues $\lambda_n(M;U) + ih\sin\theta\,
\sigma_{3\, ab}$. The infinite-volume limit of the trace thus
becomes
\begin{eqnarray}
  \lim_{V_4\to\infty}\frac{1}{V_4}{\sf Tr}\,
  \frac{i}{\gamma_5W(m_0) + h \gamma_5
  \exp\left\{i\theta\gamma_5\sigma_{3\, ab}\right\}} 
  =\int_{-\infty}^\infty d\lambda\, \rho_U(\lambda;M)\,
  \frac{i}{\lambda + i h\sin\theta\,\sigma_{3\, ab}} \ .
\label{lativol}
\end{eqnarray}

Recall that in the continuum, the eigenstates of $-i\dirac$ 
with opposite-sign eigenvalues were connected by $\gamma_5$,
and this rendered the spectral density $\tilde\rho_A(\lambda)$
symmetric. This is not true on the lattice,
and the spectral density $\rho_U(\lambda;m_0)$ 
is not symmetric in $\lambda$. Nonetheless, if $U'$ is
the parity conjugate of the configuration $U$, then 
$\rho_{U^\prime}(\lambda;m_0)=\rho_U(-\lambda;m_0)$\cite{ol}.
We can thus form a symmetric spectral density by averaging over
parity conjugated configurations, which we implicitly assume 
has been done throughout the following.

All that remains now is to take the $h\to0^+$ limit. At non-zero
lattice spacing, the spectral density \hbox{$\rho_U(\lambda;m_0+
h\cos\theta)$ uniformly converges to $\rho_U(\lambda;m_0)$} as
$h$ vanishes, and therefore the limit may be brought inside the
integral and allowed to act separately upon each term in the
integrand.\footnote{In the continuum, there is a problem 
with uniform convergence for massless quarks, but this will dealt
with at the end of this section.} As $\sin\theta$ is strictly
positive, it may be absorbed into a redefinition of $h$ in 
the denominator of (\ref{lativol}), and therefore (\ref{delpp})
gives the $\theta$-independent result \hbox{$\langle\bar\psi_b
i\gamma_5 \psi_a \rangle_U = -\pi\, \rho_U(0;m_0)
\sigma_{3\,ab}$}. Remembering that the condensate vanishes when
$\theta=0$, and upon rotating to an arbitrary flavor direction
specified by $\hat{\lpmb \theta}$, we therefore find
\begin{equation}
  \langle \bar\psi_b i\gamma_5 \psi_a \rangle_U =
  \left\{ \begin{array}{cc}
  -  \pi\, \rho_U(0;m_0)\,  
  [\hat{\lpmb \theta}   \cdot {\lpmb \sigma}]_{ab} & 
  ~~~ \theta > 0 \\
  0 & ~~~ \theta=0 \end{array} \right. \ ,
\label{latnonfdiag}
\end{equation}
with the gauge-averaged condensates $\langle \bar\psi_b 
i\gamma_5 \psi_a \rangle$ taking the same form, except that 
one uses the gauge-averaged spectral density
\begin{equation}
  \rho(\lambda;m_0) = 
  \int {\cal D}U \exp\left\{-S_g[U]\right\}\,
  \left[{\rm det}\, \gamma_5 W(m_0)\right]^2 \,
  \rho_U(\lambda;m_0) \ .
\end{equation}
Equation (\ref{condlattwo}) immediately follows.
It can be shown that the gap closes, thereby rendering
$\rho_U(0;m_0)$ non-zero, only when the dimensionless
bare mass $m_0$ is negative\cite{ol}.

While the calculational techniques presented in this section 
are not the most natural for the continuum, one could nonetheless
follow the same formal steps that led up to (\ref{lativol}). 
If the $h\to0^+$ limit can be taken inside the integral, then
for quarks of physical mass $m$ (and for non-zero $\theta$), one
obtains the $\theta$-independent continuum result \hbox{ $\langle
\bar\psi_b i\gamma_5 \psi_a \rangle_A =-\pi
\rho_A(0;m)[\hat{\lpmb \theta} \cdot{\lpmb\sigma}]_{ab}$}, with
$\rho_A(\lambda;m)$ being the spectral density for the Hermitian
operator $\gamma_5(\dirac+m)$. When $m$ is
non-zero, we see from (\ref{contsd}) that $\rho_A(0;m)=0$,
and the condensate therefore vanishes. This agrees with with
(\ref{condmb}), which was obtained using the methods of the
previous section. However, if $m$ vanishes from the start, 
from (\ref{contsd}) it is apparent that even though
$\rho_A(\lambda;h)$ converges to $\rho_A(\lambda;0)$ as $h\to0$,
it does not do so uniformly. Thus, we may not interchange the
spectral integral and the $h\to0^+$ limit in (\ref{lativol}), and
the calculational techniques of this section break down. In this
case, the methods of the previous section must be used and
(\ref{nonfdiagb}) is the correct expression for the condensate. 

\end{appendix}

\end{document}